Notice: This manuscript has been authored by UT-Battelle, LLC, under Contract No. DE-AC0500OR22725 with the U.S. Department of Energy. The United States Government retains and the publisher, by accepting the article for publication, acknowledges that the United States Government retains a non-exclusive, paid-up, irrevocable, world-wide license to publish or reproduce the published form of this manuscript, or allow others to do so, for the United States Government purposes. The Department of Energy will provide public access to these results of federally sponsored research in accordance with the DOE Public Access Plan (http://energy.gov/downloads/doe-public-access-plan).

# Physics makes the difference: Bayesian optimization and active learning via augmented Gaussian process

Maxim A. Ziatdinov,[1,2] Ayana Ghosh,[1,2] and Sergei V. Kalinin[1]

[1] Center for Nanophase Materials Sciences, Oak Ridge National Laboratory, Oak Ridge, TN 37831 USA

[2] Computational Sciences and Engineering Division, Oak Ridge National Laboratory, Oak Ridge, TN 37831 USA

Both experimental and computational methods for the exploration of structure, functionality, and properties of materials often necessitate the search across broad parameter spaces to discover optimal experimental conditions and regions of interest in the image space or parameter space of computational models. The direct grid search of the parameter space tends to be extremely time-consuming, leading to the development of strategies balancing exploration of unknown parameter spaces and exploitation towards required performance metrics. However, classical Bayesian optimization strategies based on Gaussian process (GP) do not readily allow for the incorporation of the known physical behaviors or past knowledge. Here we explore a hybrid optimization/exploration algorithm created by augmenting the standard GP with a structured probabilistic model of the expected system's behavior. This approach balances the flexibility of the non-parametric GP approach with a rigid structure of physical knowledge encoded into the parametric model. The fully Bayesian treatment of the latter allows additional control over the optimization via the selection of priors for the model parameters. The method is demonstrated for a noisy version of the classical objective function used to evaluate optimization algorithms and further extended to physical lattice models. This methodology is expected to be universally suitable for injecting prior knowledge in the form of physical models and past data in the Bayesian optimization framework.

Modern scientific research is based on the synergy of experimental and theoretical methods for the exploration and prediction of states of matter. Computational methods[1-6] ranging from lattice models and classical molecular dynamics to advanced density functional and quantum Monte Carlo techniques provide wealth of information on thermodynamic, dynamic, and electronic properties of materials. Electron and scanning probe microscopies[7, 8] and neutron and X-ray scattering methods[9, 10] provide huge volumes of structural and functional data. Finally, combinatorial, and automated synthesis and high-throughput characterization[11-13] now allows rapid exploration of multidimensional compositional spaces for complex functional materials.

Common for the experimental and computational methods alike is the need for search across broad parameter spaces. In theory materials prediction necessitates exploration of large chemical spaces of molecules encoded via SMILEs, SELFIES, or other descriptors, or compositional spaces of complex materials. Very similar challenges exist in the experimental domains ranging from automated synthesis to imaging and spectroscopy. For example, tuning of physical instrument and search of the region of interest are both searches in parameter space, to find optimal experimental conditions and regions of interest within material respectively. However, the direct grid search of the parameter space tends to be extremely time-consuming, leading to the development of strategies balancing exploration of unknown parameter spaces and exploitation towards required performance metrics. The classical approach for the this is the Gaussian process (GP) based Bayesian optimization (BO).[14] This method balance the learning of the correlations in the parameter space (via kernel function) with the exploration balancing the uncertainty and maximization of a certain target function combined in a single acquisition function.

However, classical GP-based BO strategies do not readily allow for the incorporation of the known physical behaviors or past knowledge. In other words, the GP-based BO methods create a fully non-parametric model of the system based on the prior observations, with the "learned" physics of the system reflected in the kernel function. As such, the latter is roughly equivalent to the parametrized via a certain functional form correlation function. At the same time, it is well-known that such representation tends to produce suboptimal results if the functional form is incorrect, and loses information contained in higher-order correlations. As such, this approach is limited if behaviors of interest change differently in different regions of parameter space, e.g. have sharp boundaries, etc. While a large number of GP/BO models allowing for the greater flexibility of kernel function, warping of the parameter space, etc. are continuously being developed,[15-17] the overall limitations persist.

Comparatively, physics applications are typically associated with a large volume of domain-specific knowledge. For example, in exploring the lattice models in statistical physics, the asymptotic behaviors of the relevant parameters in the vicinity of phase transition and in the large/small temperature limits are generally known, but not the transition temperature or universality classes. Similarly, in materials discovery, the evolution of properties of interest such as photoluminescent peak position or stability are strongly correlated with the composition-dependent band gap or thermodynamic stability. In ferroelectric materials, the thickness dependence of the coercive field is defined by the phenomenological Kay-Dunn laws. Virtually in all areas of physical sciences, past knowledge is encoded in the form of symbolic models. In some cases, the models can be derived from the underpinning microscopic models or follow from the conservation laws; in many others, they represent the phenomenological behavior of the system. Nonetheless, this knowledge is generally not incorporated in the Bayesian optimization framework.

Here we explore a hybrid optimization/exploration algorithm created by augmenting the standard GP with a structured probabilistic model of the expected system's behavior. This approach balances the flexibility of the non-parametric GP approach with a rigid structure of physical knowledge encoded into the parametric model. The method is demonstrated for a noisy version of the classical objective function used to evaluate optimization algorithms and further extended to physical lattice models and is expected to be universally suitable for injecting prior knowledge in the form of physical models and past data into the Bayesian optimization framework.

*Gaussian process*

Given the dataset $D = \{x^i, y^i\}_{i=1,\dots,N}$, where $x$ and $y$ are input features and output targets (referred to as 'training' data in the ML community), respectively, the GP model can be defined as

$$\mathbf{y} \sim MVNormal(\mathbf{m}, \mathbf{K}) \tag{1a}$$

$$K_{ij} = \sigma^2 \exp(0.5(x_i - x_j)^2/l^2) \tag{1b}$$

$$\sigma \sim LogNormal(0, s_1) \tag{1c}$$

$$l \sim LogNormal(0, s_2) \tag{1d}$$

where *MVNormal* stands for multivariate normal distribution, **m** is a mean function which is usually chosen to be constant, and **K** is a covariance function (kernel), for which we chose a standard radial basis function with output scale $\sigma$ and length scale $l$. We also assume there is normally distributed observation noise, $\boldsymbol{\varepsilon} \sim Normal(0, \sigma_{noise}\mathbf{I})$, such that $\mathbf{y}_{\text{noisy}} = \mathbf{y} + \boldsymbol{\varepsilon}$. Implementation wise, this noise is absorbed into the covariance function $K$. To get posterior samples for the GP model parameters, we use a Hamiltonian Monte Carlo (HMC)[18] algorithm. After inferring the parameters of GP model, we can use it to obtain the expected function values and associated uncertainty on new, previously not seen by the model, data (the so-called 'test' data). Specifically, we sample from the multivariate normal posterior over the model outputs at the provided points $X_*$:

$$\mathbf{f}_*^i \sim MVNormal\left(\mu_{\boldsymbol{\theta}^i}^{\text{post}}, \Sigma_{\boldsymbol{\theta}^i}^{\text{post}}\right) \tag{2a}$$

$$\mu_{\boldsymbol{\theta}^i}^{\text{post}} = \mathbf{m}(X_*) + \mathbf{K}(X_*, X|\boldsymbol{\theta}^i)\mathbf{K}(X, X|\boldsymbol{\theta}^i)^{-1}(\mathbf{y} - \mathbf{m}(X)) \tag{2b}$$

$$\Sigma_{\boldsymbol{\theta}^i}^{\text{post}} = \mathbf{K}(X_*, X_*|\boldsymbol{\theta}^i) - \mathbf{K}(X_*, X|\boldsymbol{\theta}^i)\mathbf{K}(X, X|\boldsymbol{\theta}^i)^{-1}\mathbf{K}(X, X_*|\boldsymbol{\theta}^i) \tag{2c}$$

where $\boldsymbol{\theta}^i = [\sigma^i, l^i, \sigma_{noise}^i]$ is a single HMC posterior sample containing kernel hyperparameters and model noise.

*Bayesian optimization*

The GP posterior in Eq. (2) can be used for selecting the next measurement/evaluation point in the active learning or Bayesian optimization setting. This is achieved by minimizing (or maximizing, depending on a particular problem) the so-called acquisition function. Perhaps the simplest (but nevertheless rather effective) version of the acquisition function is a Thompson sampler, which represents a single draw, $\mathbf{f}_*^i$, from posterior samples. Another acquisition function, known as the upper confidence bound or UCB, is derived from a linear combination of the predicted mean function value, $\overline{\mathbf{f}_*}$, and associated variance ('uncertainty'), $\mathbb{V}[\mathbf{f}_*]$, across the sampled predictions. Finally, in a pure exploratory regime referred to as kriging,[19] the selection of the next points is guided by the minimization of uncertainty $\mathbb{V}[\mathbf{f}_*]$.

Hence, a single Bayesian optimization (BO) step consists of i) obtaining/updating the GP posterior over the model outputs at the provided points $X_*$ given the sparse measurements *D*, ii) deriving the acquisition function, iii) selecting the next measurement point based on the minimum (or maximum) value of the acquisition function, iv) performing measurement in the 'suggested' point and adding the measured value to the dataset *D*. The goal of Bayesian optimization is usually to quickly identify regions where a particular behavior is maximized or minimized. As such, it has been actively used in many domains, ranging from organic synthesis[20] to hyperparameter optimization in deep learning.[21]

The limitation of the standard GP-based BO is that it does not readily allow for the incorporation of the known physical behaviors or past knowledge. This means that no domain-specific information or theoretical insights are factored in the selection of the next query point(s), potentially resulting in a suboptimal sampling of the parameter spaces of physical systems. In other words, whereas GP-BO is an optimal exploration strategy in the absence of prior knowledge, this is not the case in most domain applications, including materials science, physics, and chemistry, where prior domain expertise and first-principles simulations are often the key factors guiding the exploration.

*Augmented GP-BO and toy model*

To overcome the limitations of the classical GP-BO we propose augmenting GP with a structured probabilistic model of the expected system's (physical) behavior. Specifically, we substitute a constant mean function **m** in Eq. (1-2) with a probabilistic model whose parameters are inferred together with the kernel parameters via the HMC. This probabilistic model reflects our prior knowledge about the system, but it does not have to be precise. In other words, the model can have a different functional form, as long as it captures the general trend in the data. In the language of machine learning, this can be interpreted as adding the so-called inductive bias[22] to our GP model. The Eq. (2b) then becomes

$$\mu_{\theta^i}^{post} = \mathbf{m}(X_*|\phi^i) + \mathbf{K}(X_*, X|\theta^i)\mathbf{K}(X, X|\theta^i)^{-1}(\mathbf{y} - \mathbf{m}(X|\phi^i)) \qquad (3)$$

where $\phi^i$ is a single HMC posterior sample with the learned model parameters. This leads to the semi-parametric Bayesian optimization algorithm that makes "physics-informed" decisions about which points in the parameter space to evaluate next. We note that our approach is different from using a deterministic GP mean function (either in analytical form[23] or as a neural network[24]) as it allows specifying a fully Bayesian probabilistic model by placing suitable priors on its parameters so that the uncertainty from the corresponding posterior samples is automatically taken into account when performing Bayesian optimization or active learning. At the same time, the flexibility of GP kernel allows us to use an imprecise and sometimes even 'incorrect' functional form, which would not be possible for a standalone probabilistic model (see Supplementary Materials). We demonstrate our approach for the toy model based on the modified Forrester function[25] and for the 1D and 2D Ising models, but it is expected to be universally suitable for injecting prior knowledge in the form of physical/domain knowledge in the Bayesian optimization framework.

We start by illustrating our idea using a modified Forrester function,[25] $f(x) = (5x - 2)^2 \sin(12x - 4)$, which is commonly used to evaluate the optimization algorithms. We note that the end-goal is to have a robust optimization algorithm capable of working with noisy experimental or simulated data. Hence, we deliberately corrupted the observations with Gaussian noise. The noisy observations and the true function are shown in Fig. 1(a). The predictive means of the standard GP model trained on the entire data in Fig. 1a along with the associated uncertainty (variance across the sampled predictions) are shown in Fig. 1(b).

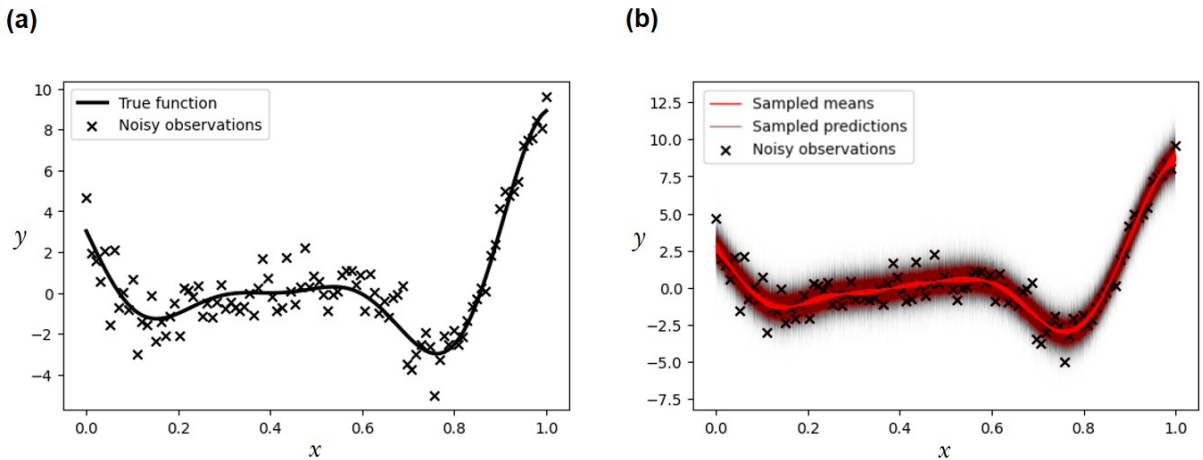

**Figure 1.** Forrester objective function. **(a)** Noisy observations of the Forrester objective function and the true function. **(b)** The posterior predictive distribution of the standard GP model trained on all the data points. The red curves show the GP predictive means according to Eq. (2b) for different HMC samples and the maroon curves show the sampled GP predictions. Here we used a 10x denser grid for predictions for better visualization of the 'uncertainty' across the GP samples.

We now proceed to the BO whose goal, in this case, is to converge to the global minima region (we cannot find the exact minimizer of the function since observations contain noise) using a minimal number of steps. First, we perform BO using the standard GP. We start with just two initial (seed) observations and run BO for 12 iterations using Thompson and UCB acquisition functions. The results are shown in Figure 2. Clearly, the BO with Thompson sampler (Fig. 2(a)) could not successfully converge to the global minima region whereas in the case of UCB (Fig. 2(b)) the optimization simply got stuck in a local minimum. The reasons for this somewhat disappointing performance are bad initialization (the seed points are near the local minimum) and relatively large observational noise. We note that for the clean data, the standard GP-BO identifies the global minimum with relative ease as shown in Supplementary Figure 1. However, the real-world measurements are always noisy. Furthermore, one usually does not have the luxury of restarting the experimental measurements (or even sometimes simulations) multiple times using different initializations.

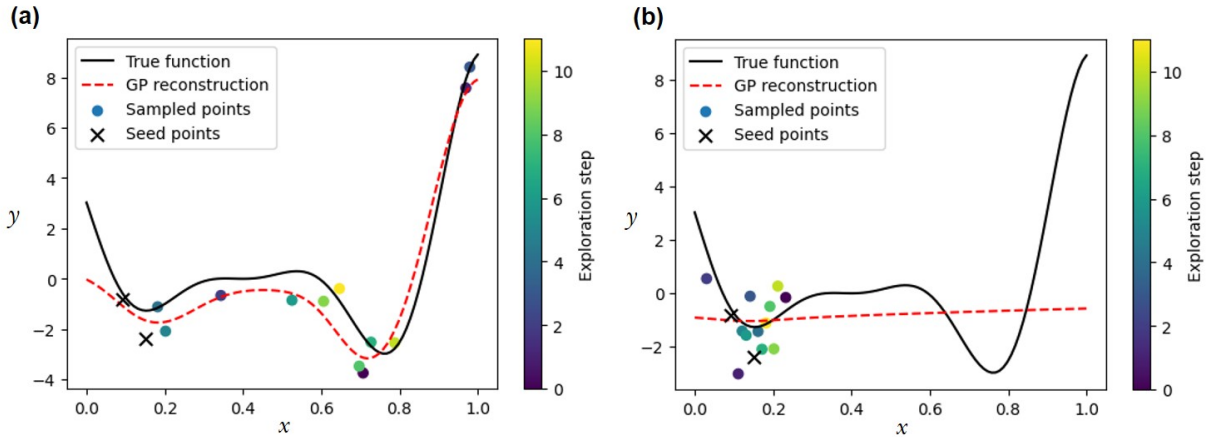

**Figure 2.** Results of the standard GP-BO for noisy observations of Forrester function. **(a)** Thompson sampling, **(b)** UCB. The "GP reconstruction" shows the center of mass of the sampled predictive means from Eq. (2b).

Next, we show how injecting prior knowledge about our objective function/data can help mitigate the aforementioned issues. Specifically, we are going to augment GP with two different structured probabilistic models. Although both models do not describe the actual objective function, they do describe certain general trends in the data. Our first probabilistic model is defined as

$$m = y_0 - \sum_{n=1}^{N} L_n \quad (N=2) \tag{4a}$$

$$y_0 \sim Uniform(-10, 10) \tag{4b}$$

$$L_n \sim \frac{A_n}{\sqrt{(x-x_n^0)^2+w_n^2}} \tag{4c}$$

$$A_n \sim LogNormal(0, 1) \tag{4d}$$

$$w_n \sim HalfNormal(.1) \tag{4e}$$

$$x_n^0 \sim Uniform(0, 1) \tag{4f}$$

This model simply tells us that there are two minima in our data but does not assume to have any prior knowledge about their relative depth (i.e., which one of them is global) and width, nor it contains information about how far apart they are. The prior draws from the model are shown in Fig. 3a. Next, we substitute the constant mean function in GP with our probabilistic model and run the BO the same way as we did earlier.

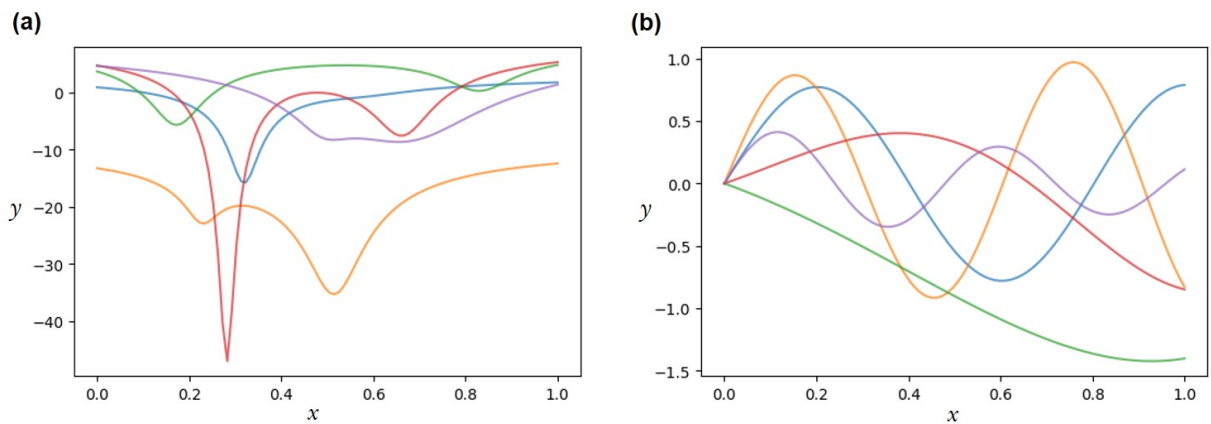

**Figure 3.** Prior predictive distributions for the two structured probabilistic models described in **(a)** Eq (4) and **(b)** Eq (5).

The results for the Thomson sampler and UCB are shown in Fig. 4 (a) and 4(b), respectively. Compared to the standard GP-BO, the GP-BO augmented with structured probabilistic model (hereafter referred to as *s*GP-BO) was able to converge to the global minima region within a relatively small number of steps. In addition, even though this was not a goal here, the *s*GP reconstruction of the objective function based on just (12+2) points closely matches the shape of the true objective function. This shows that augmenting GP with a (imprecise) model of systems behavior that captures general trend in data is enough to significantly improve the efficiency of the optimization with noisy observations. We can also inspect the statistics of posterior samples for the mean GP function after the 12 steps of *s*GP-BO. The inferred center ($x_0$) of the deeper minimum is at $0.77 \pm 0.03$, which is close to the true minimum of the objective function (0.757). We note that the optimization efficiency can be further improved by using more informative priors in Eq. (4b-f) with appropriately chosen acquisition functions. For example, the more informative priors about the minima locations can improve a search efficiency when using the acquisition functions that prioritize exploitation over exploration (see Supplementary Figure 2).

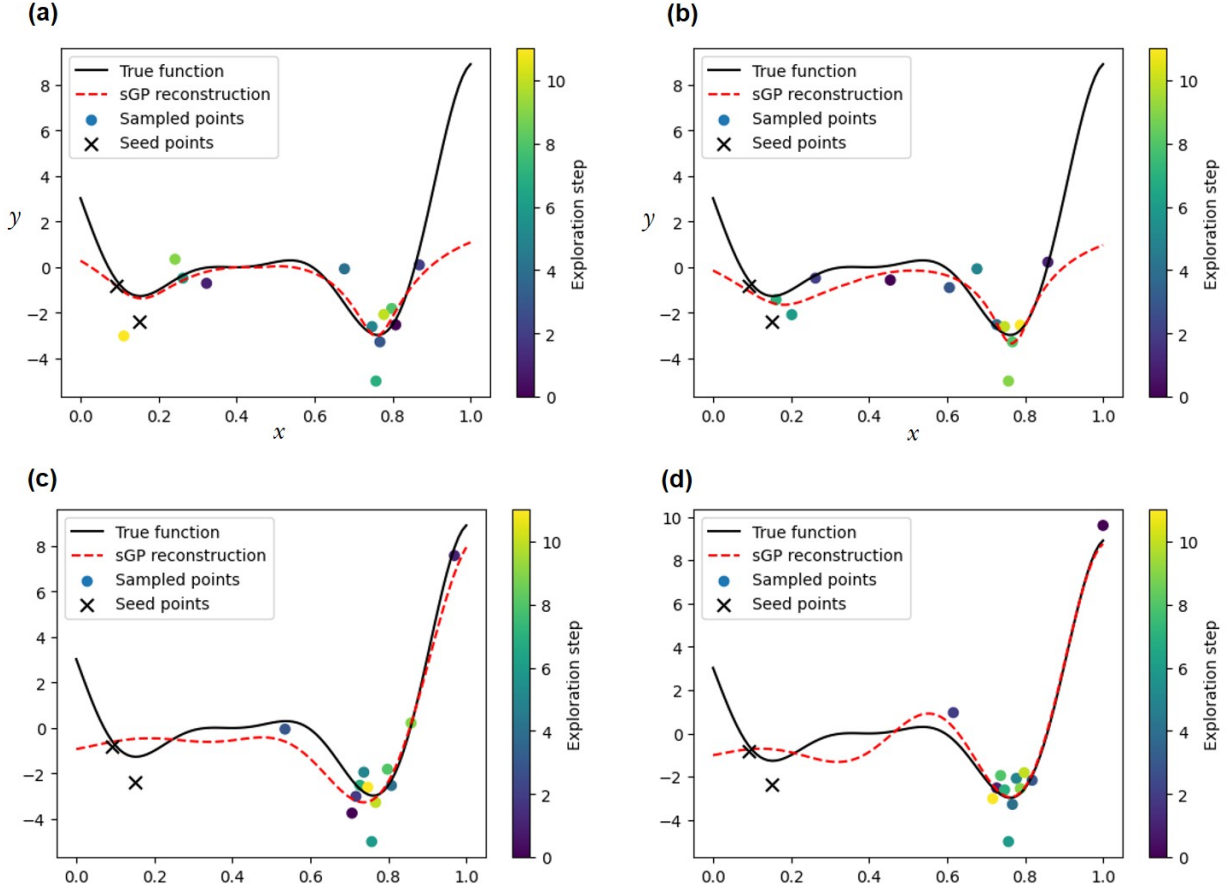

**Figure 4.** Results of the structured GP-BO for the noisy observations of Forrester function with a probabilistic model from Eq. (4) **(a, b)** and Eq. (5) **(c, d)** as GP's mean function. **(a, c)** show results for Thompson sampler and **(b, d)** show results for UCB. The "sGP reconstruction" shows the center of mass of the sampled predictive means from Eq. (3).

For our second probabilistic model, we have chosen a model that represents a 'wrong' function but still *partially* captures trends in the data such as the presence of more than one minimum. The second model is defined as

$$m = A\mathrm{e}^{ax}\sin(bx) \quad (5a)$$

$$A \sim LogNormal(0, 1) \quad (5b)$$

$$a \sim Normal(1, 2) \quad (5c)$$

$$b \sim Normal(10, 5) \qquad (5d)$$

The prior draws for the second model are shown in Fig. 3(b) and the *s*GP-BO results for the Thomson sampler and UCB are shown in Fig. 4 (c, d). One can see that while the quality of the overall reconstruction has suffered, we were able nevertheless to converge on the global minimum region. We note that this is only possible because we use a hybrid of the GP and probabilistic model and would not be possible if we use only the latter (see Supplementary Figure 3).

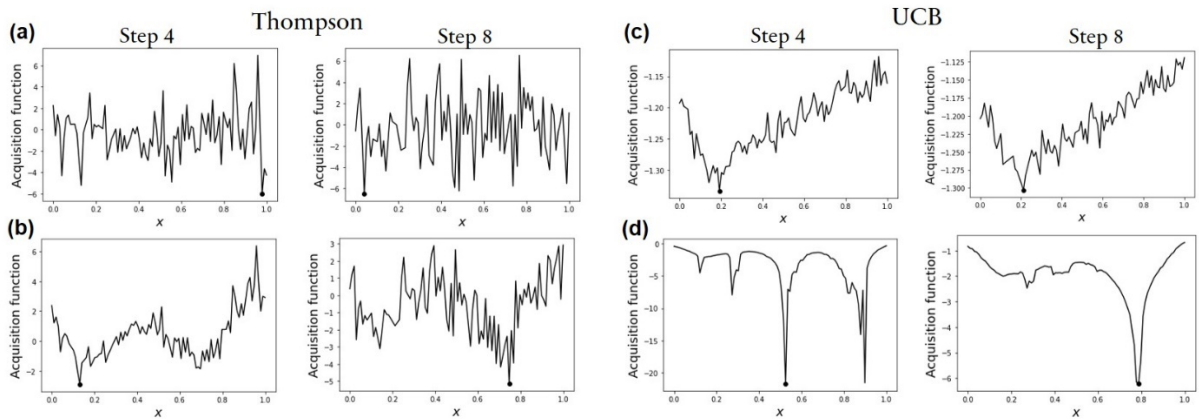

**Figure 5.** Acquisition functions at different optimization steps. **(a, b)** Thompson acquisition function for the **(a)** GP-BO and **(b)** *s*GP-BO. **(c, d)** UCB acquisition function for the **(c)** GP-BO and **(d)** *s*GP-BO

To better understand the difference between the standard GP-BO and the GP-BO augmented with a structured probabilistic model, we compared the acquisition functions for both algorithms at different steps of the optimization. In Figure 5 (a, b), we show the Thompson acquisition function at optimization steps 4 and 8 for the standard GP-BO (Fig. 5(a)) and the GP-BO augmented by the probabilistic model in Eq. (4) (Fig. 5(b)). For the standard GP-BO, the acquisition function is simply too noisy to guide the optimization algorithm. On the other hand, in the case of the *s*GP-BO, the acquisition function starts reflecting the general trends in the data already at the early steps of the optimization. In the case of the UCB acquisition function (Fig. 5(c, d)), it remains largely unchanged for the standard GP-BO (Fig. 5(c)) as the algorithm 'assumes' that it already found a global minimum. At the same time, for the *s*GP-BO (Fig. 5(d)), the injection of prior knowledge (that there is more than one peak) helps the algorithm to climb out of the local minimum and converge on the location of the global minimum.

To ensure the reproducibility of our results, we performed a systematic comparison of two GP-BO approaches for 20 different initializations of initial observations (using different pseudo-random number generator seeds). We introduced the efficiency (*e*) parameter that defines the number of points discovered by the algorithm in the ±0.05 vicinity of the true minimum after 10 steps divided by the total number of points lying in the same region for a given discretization of the X space. The ideal value is $e = 1$. The results shown in Fig. 6 show that our approach clearly outperforms the standard GP-BO for both Thompson and UCB acquisition functions.

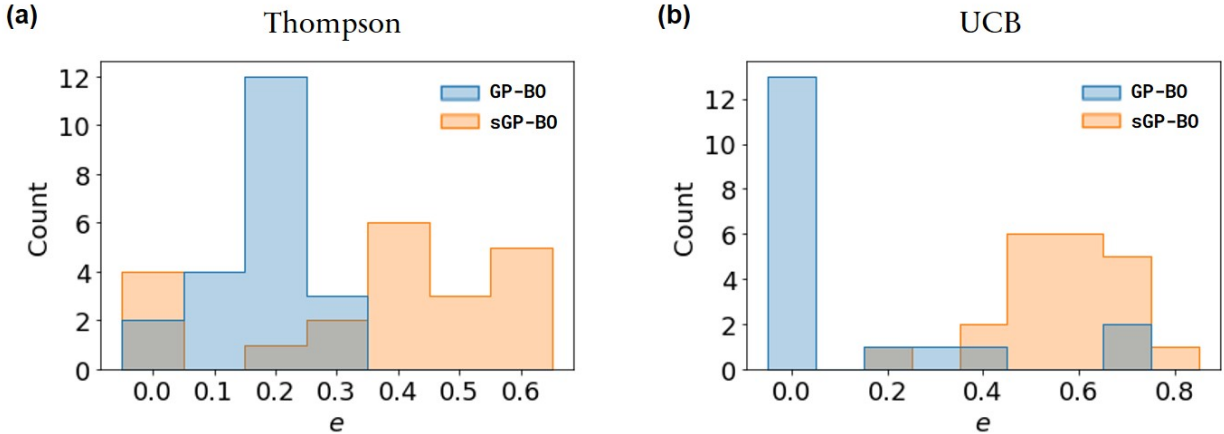

**Figure 6.** Efficiency of the two different BO approaches for **(a)** Thomson sampler and **(b)** UCB acquisition function. The efficiency ($e$) is calculated as a ratio of the points sampled by an optimization algorithm within ±0.05 of the true minimum to the total number of points lying within the same range for a given discretization of the X space ($\Delta x=0.01$). For the 10 optimization steps used here, the highest possible efficiency is 1.

*Physical systems*

Having verified that our algorithm works on synthetic data, we next apply it to the 1D and 2D Ising model. The Hamiltonian of the Ising Model with nearest-neighbor interactions can be written as $\mathcal{H} = -\sum_{<ij>} J_{ij} \mathbf{S}_i \cdot \mathbf{S}_j$ where $J$ represents the spin-spin interaction term and $\mathbf{S}_i$ are the individual spins on each of the lattice sites. The parameters of the Ising model simulations are the same as in Ref [26]. Here, we are going to use the classical thermodynamic properties such as susceptibility and magnetization for our objective functions to compare the two GP-BO approaches.

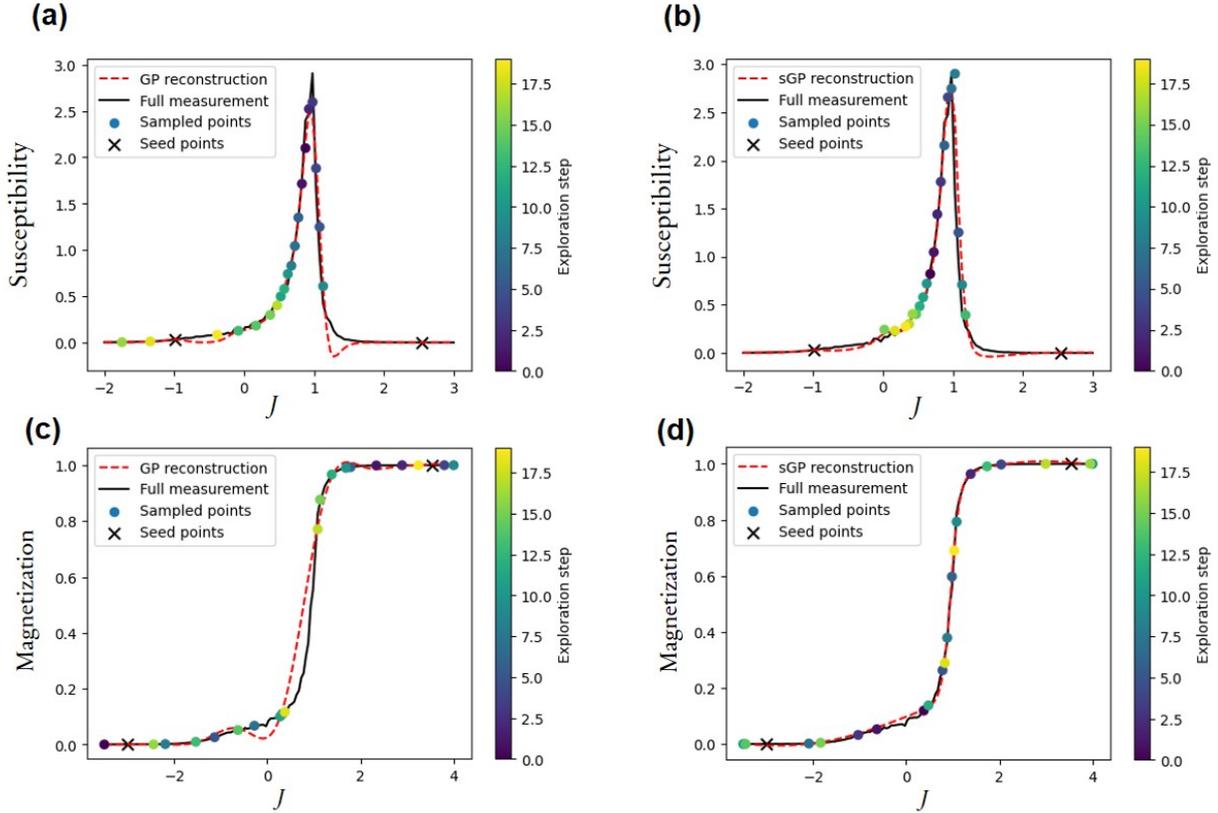

**Figure 7.** BO and active learning for 1D Ising model. **(a, b)** Optimization of the $J$ parameter to maximize the susceptibility using UCB acquisition function with **(a)** standard GP-BO and **(b)** $s$GP-BO. **(c, d)** Kriging-based exploration of the magnetization across the phase transition with **(c)** standard GP kriging and **(d)** $s$GP kriging.

We start by using BO to identify values of the $J$ parameter that maximize the susceptibility in the 1D Ising model. For the $s$GP-BO, we use a simple Gaussian peak model of the form $A e^{(-(J-J_0)^2/w^2)}$ with (log-)normal priors on its parameters as our GP mean function. The results for classical GP-BO and $s$GP-BO are shown in Fig. 7(a, b). One can see that while results are somewhat close (due to the relatively low noise and presence of only one peak) the incorporation of prior knowledge allows avoiding unphysical behaviors, such as at $J \approx 1.2$ in Fig. 7 (a), when only a limited number of observations is available.

For the magnetization, which does not have a global minimum/maximum, we used a pure uncertainty-based exploration (kriging). Here, the goal is to explore and reconstruct its behavior in two different phases and, most importantly, at the phase boundary. The results for the 1D case are shown in Figure 7(c, d). For the $s$GP, we used a logistic function of the form $A/\tanh\left(\frac{J-J_0}{w}\right)$, with a uniform prior on $J_0$ and log-normal priors on $A$ and $w$, as our structured probabilistic model. Clearly, the incorporation of prior knowledge about the system's behavior allowed for exploration of the transition region and better overall reconstruction (Fig. 7d). For the 2D case, we used a probabilistic model of the form $A/\tanh\left(\frac{f(J_1)+f(J_2)}{w}\right)$ where $f(J)$ is a third-degree polynomial with normal priors on its parameters. Note that while this is not the actual functional form of the phase transition in the system, it allows (with a proper choice of priors) to incorporate our knowledge that this phase transition can be in general described by a sigmoid-like structure but the actual phase boundary is a curve with a potentially complex shape. This should allow for a sufficient flexibility needed to map and reconstruct the phase boundary using a small number of measurements. The results for the standard GP kriging is shown in Figure 8 (a). Evidently, it could not localize the phase boundary and provided an overall poor reconstruction (Fig. 8 (b)). In comparison, the *s*GP kriging was able to map the phase boundary (Fig. 8 (b)) and provide a reconstruction of a sufficient quality (Fig. 8 (d)) within the same number of exploration steps.

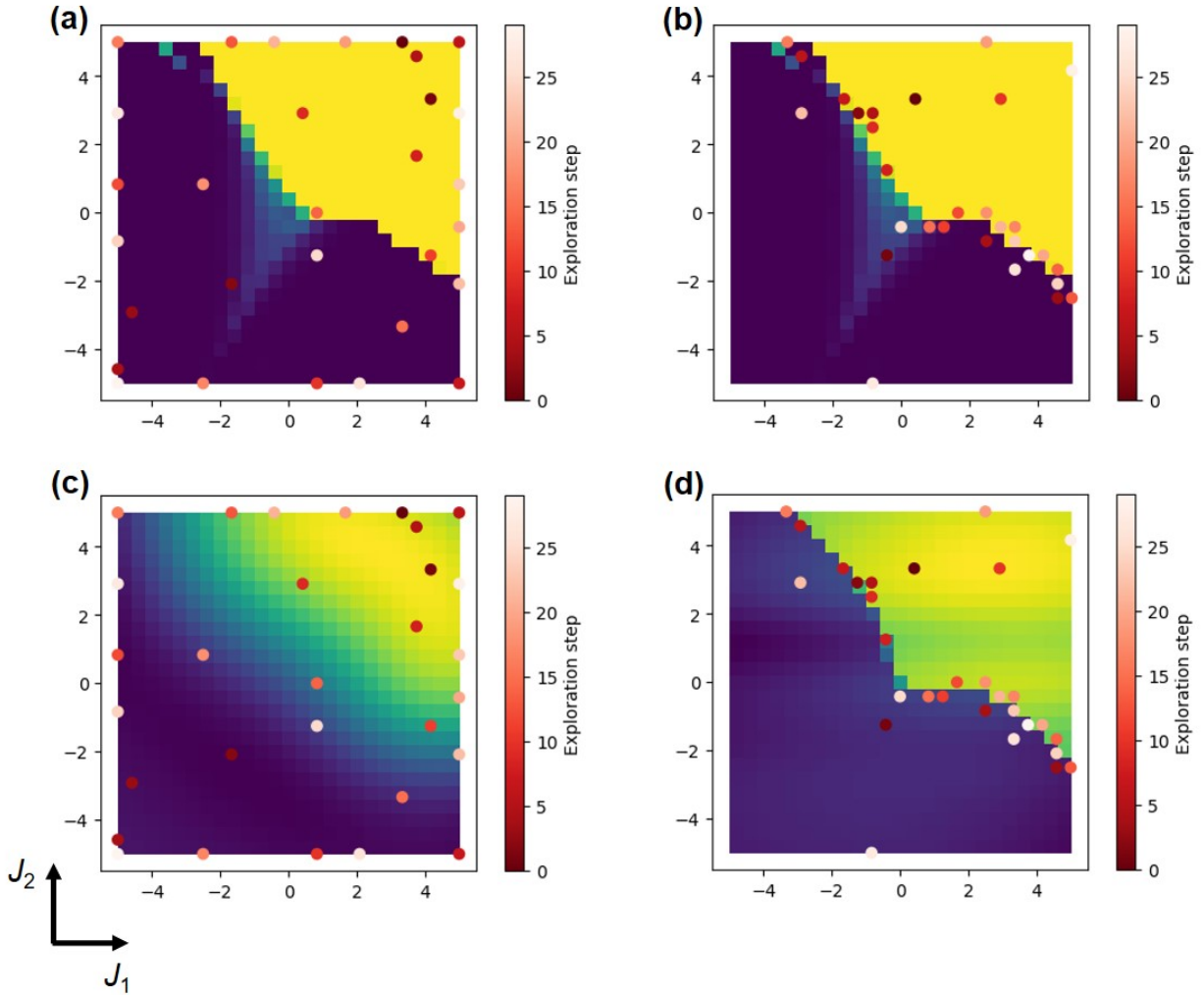

**Figure 8.** Active learning of magnetization behavior in 2D Ising model. **(a, b)** Points sampled by **(a)** standard GP kriging and **(b)** *s*GP kriging overlaid with the fully measured system ('ground truth'). **(c, d)** Data reconstructed from the points sampled using the **(c)** standard GP and **(d)** *s*GP.

In summary, we have demonstrated how the augmentation of Gaussian process with a (fully Bayesian) probabilistic model of expected system's physical behavior allows for a more efficient optimization and active learning of system's properties. This was demonstrated for noisy observations of the standard objective function used to evaluate optimization algorithms and for the 1D and 2D Ising model in physics. The current approach can be extended to more complex physical systems where in the absence of any prior knowledge one can start with a standard GP and after sufficient number of observations (allowing to come up with a possible model of system's behavior) adds a structured probabilistic model, although a more sophisticated interplay between GP and *s*GP is possible. Finally, in addition to incorporating our knowledge about possible functional forms, one can also incorporate a partial knowledge of causal links in a system potentially allowing for an even more efficient optimization and active learning.

**Acknowledgements:**

This work was supported (M.Z.) and performed at Oak Ridge National Laboratory's Center for Nanophase Materials Sciences (CNMS), a U.S. Department of Energy, Office of Science User Facility, and supported by the Energy Frontier Research Centers program: CSSAS–The Center for the Science of Synthesis Across Scales–under Award Number DE-SC0019288, located at the University of Washington (S.V.K., A.G.).

**Methods:**

The GP-BO and sGP-BO routines were implemented in JAX[27] using the iterative No-U-Turn-Sampler[28, 29] for HMC. The UCB acquisition function was defined as $\bar{\mathbf{f}}_* + k\sqrt{\mathbb{V}[\mathbf{f}_*]}$, with $k = -0.5$ for minimizing the Forrester objective function and $k = 0.5$ for maximizing the susceptibility in Ising model. The code will be made available at
https://github.com/ziatdinovmax/AugmentedGaussianProcess .

# Supplemenatry Material

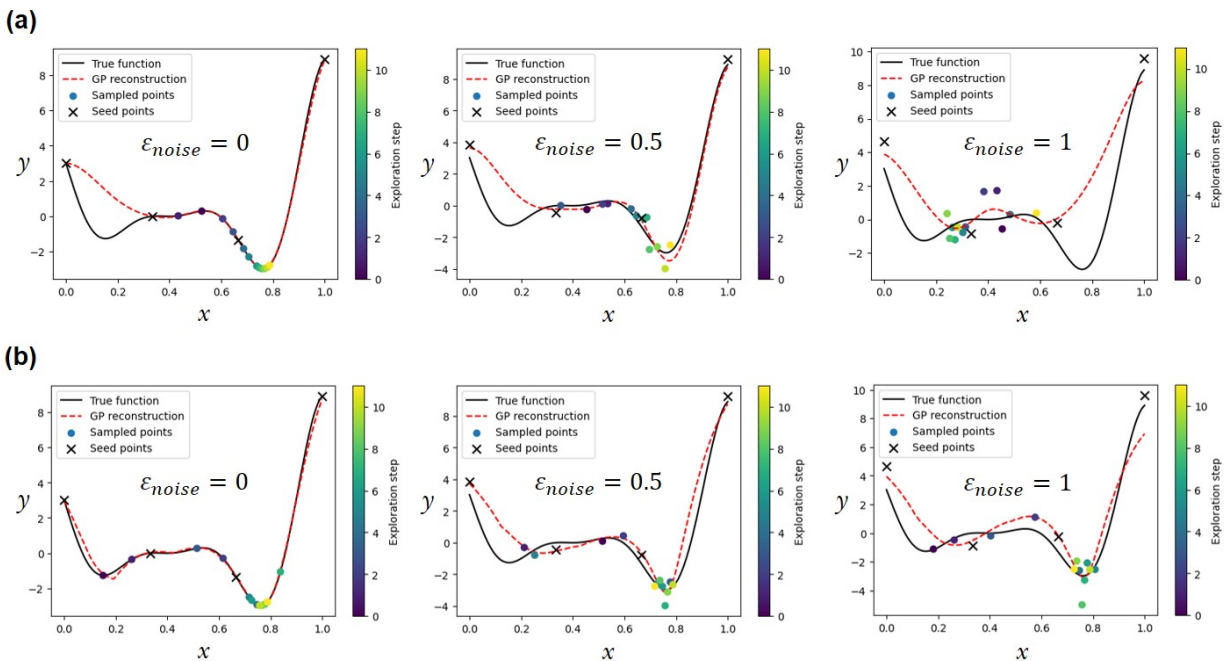

**Supplementary Figure 1.** Performance of **(a)** standard GP-BO and **(b)** *s*GP-BO (augmented with probabilistic model from Eq. 4) for different levels of noise with the UCB acquisition function. The $\varepsilon_{noise}$=1 corresponds to the noise level used in the main text (see Fig. 1). Note that here we initialized seed points uniformly.

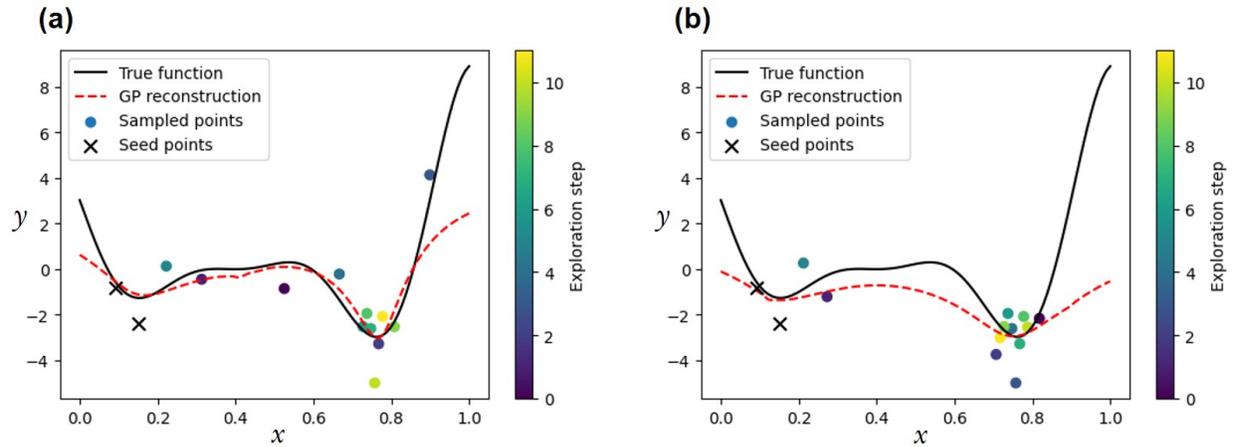

**Supplementary Figure 2.** Comparison of *s*GP-BO with the expected improvement (EI)[1] acqusition function using (a) the same priors as in Eq. (4) and (b) slightly more informative priors where the positions ($x_0$) are sampled from the (0.1, 0.3) and (0.7, 0.9) uniform distributions. In the second case, the algorithm shows a better efficiency in identifying the global minimum region even though the overall reconstruction quality has suffered as a result of it (the reconstruction was not the goal here, however).

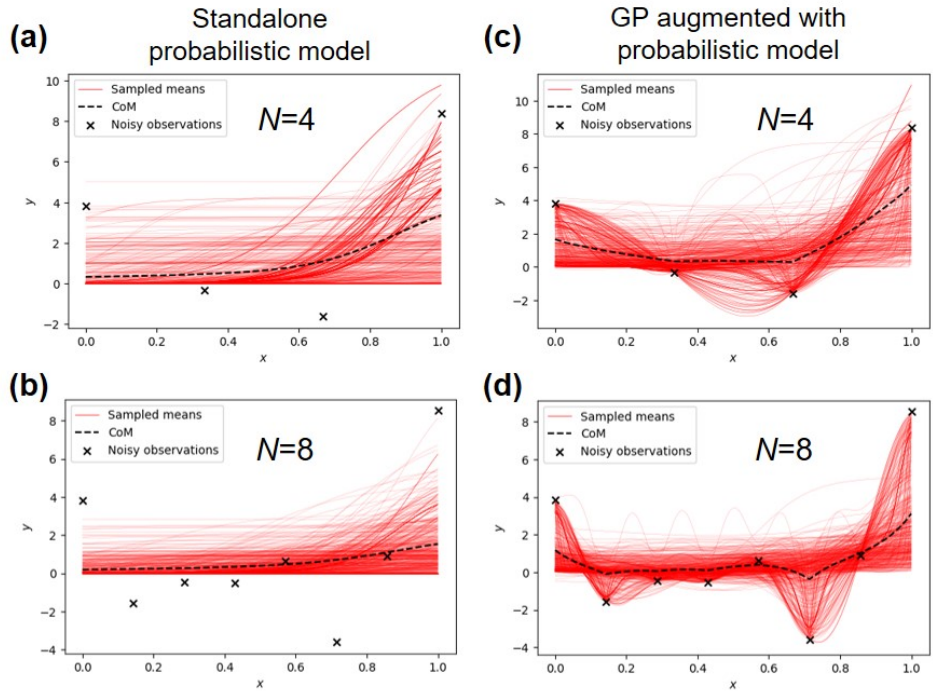

**Supplementary Figure 3.** Example of using incorrect probabilistic model (does not capture any trend in data) for obtaining predictive posterior distributions given $N$ noisy observations of the Forrester function. **(a, b)** Standalone probabilistic model of the form $\frac{A}{(1+e^{-\beta(x-x_0)})}$ with normal priors on its parameters and **(c, d)** GP augmented by the probabilsitic model of the same form with the exact same priors. "CoM" stands for the center of mass of the sampled means.